Coherent Ising machines - Quantum optics and neural network perspectives -

Y. Yamamoto[1], T. Leleu[2,3], S. Ganguli[4], and H. Mabuchi[4]

1. Physics & Informatics Laboratories, NTT Research, Inc., 1950 University Ave. #600, East Palo Alto, CA 94303, USA
2. Institute of Industrial Science, The University of Tokyo, 4-6-1 Komaba, Meguro-ku, Tokyo 153-8505, Japan
3. International Research Center for Neurointelligence, The University of Tokyo, 7-3-1 Hongo Bunkyo-ku, Tokyo 113-0033, JAPAN
4. Department of Applied Physics, Stanford University, Stanford, CA94305, USA

Author to whom correspondence should be addressed: yoshihisa.yamamoto@ntt-research.com

ABSTRACT

A coherent Ising machine (CIM) is a network of optical parametric oscillators (OPOs), in which the "strongest" collective mode of oscillation at well above threshold corresponds to an optimum solution of a given Ising problem. When a pump rate or network coupling rate is increased from below to above threshold, however, the eigenvectors with a smallest eigenvalue of Ising coupling matrix $[J_{ij}]$ appear near threshold and impede the machine to relax to true ground states. Two complementary approaches to attack this problem are described here. One approach is to utilize squeezed/anti-squeezed vacuum noise of OPOs below threshold to produce coherent spreading over numerous local minima via quantum noise correlation, which could enable the machine to access either true ground states or excited states with eigen-energies close enough to that of ground states above threshold. The other approach is to implement real-time error correction feedback loop so that the machine migrates from one local minimum to another during an explorative search for ground states. Finally, a set of qualitative analogies connecting the CIM and traditional computer science techniques are pointed out. In particular, belief propagation and survey propagation used in combinatorial optimization are touched upon.

Introduction

Recently, various heuristics and hardware platforms have been proposed and demonstrated to solve hard combinatorial or continuous optimization problems. The cost functions to be minimized in those problems are either Ising Hamiltonian, $\mathcal{H}_{\text{Ising}} = \sum J_{ij}\sigma_i\sigma_j$, for binary variables $\sigma_i = \pm 1$ or XY Hamiltonian, $\mathcal{H}_{\text{XY}} = \sum J_{ij}\cos(\theta_i - \theta_j)$, for continuous variables $\theta_i = [0, 2\pi]$, which is mapped to the energy landscape of classical spins,[1][2][3] quantum spins,[4][5] solid state devices[6][7][8] or neural networks.[9][10] Convergence to a ground state is assured for a slow enough decrease of the temperature in simulated annealing.[11] An alternative approach based on networks of optical parametric oscillators (OPOs)[12][13][14][15][16][17][18] and Bose-Einstein condensates[19][20] has been also actively pursued, in which the target function is mapped to a loss landscape. Intuitively, by increasing the gain of such an open-dissipative network with a slow enough speed by ramping an external pump source, a lowest-loss ground state is expected to emerge as a single oscillation/condensation mode.[13][21] In practice, ramping the gain of such a system results in a complex series of bifurcations that may guide or divert evolution towards optimal solution states.

One of the unique theoretical advantages of the second approach, for instance in a coherent Ising machine (CIM),[12][13][14][15][16] is that quantum noise correlation formed among OPOs below oscillation threshold could in principle facilitate quantum parallel search across multiple regions of phase space.[22] Another unique advantage is that following the oscillation-threshold transition, exponential amplification of the amplitude of a selected ground state is realized in a relatively short time scale of the order of a photon lifetime. In a non-dissipative degenerate parametric oscillator, two stable states at above bifurcation point co-exist as a linear superposition state.[23][24] On the other hand, the network of dissipative OPOs[13][14][15][16][17] changes its character from a quantum analog device below threshold to a classical digital device above threshold. Such quantum-to-classical crossover behavior of CIM guarantees a robust classical output as a computational result, which is in sharp contrast to a standard quantum computer based on linear amplitude amplification realized by Grover algorithm and projective measurement.[25]

A CIM based on coupled OPOs, however, has one serious drawback as an engine for solving combinatorial optimization problems: mapping of a cost function to the network loss landscape often fails due to the fundamentally analog nature of the constituent spins, *i.e.*, the possibility for constituent OPOs to oscillate with unequal amplitudes. This problem is particularly serious for a frustrated spin model. The network may spontaneously find an excited state of the target Hamiltonian with lower effective loss than a true ground state by self-adjusting oscillator amplitudes.[13] An oscillator configuration with frustration and thus higher loss may retain only small probability amplitude, while an oscillator configuration with no frustration and thus smaller loss acquires a large probability amplitude. In this way, an excited state can achieve a smaller overall loss than a ground state (see Fig. 6 of [13]). Recently, the use of an error detection and correction feedback loop has been proposed to suppress this amplitude heterogeneity problem[20] and the improved performance of such a feedback controlled CIM has been numerically confirmed.[26] The proposed system has a recurrent neural network configuration with asymmetric weights ($J_{ij} \neq J_{ji}$) so that it is not a simple gradient-descent system any more. The machine can escape from a local minimum by a diverging error correction field and migrate from one local minimum to another. The ground state can be identified during such a random exploration of the machine.

In this letter, we present several complementary perspectives for this computing machine, which are based on diverse, interdisciplinary viewpoints spanning quantum optics, neural networks and message passing. Along the way we will touch upon connections between the CIM and foundational concepts spanning the fields of statistical physics, mathematics, and computer science, including dynamical systems theory, bifurcation theory, chaos, spin glasses, belief propagation and survey propagation. We hope the bridges we build in this article between such diverse fields will provide the inspiration for future directions of interdisciplinary research that can benefit from the cross-pollination of ideas across multifaceted classical, quantum and neural approaches to combinatorial optimization.

Optimization dynamics in continuous variable space

CIM studies today could well be characterized as experimentally-driven computer science, much like contemporary deep learning research and in contrast to the current scenario of mainstream quantum computing. Large-scale measurement feedback coupling coherent Ising machine (MFB-CIM) prototypes constructed by NTT Basic Research Laboratories[15] are reaching intriguing levels of computational performance that, in a fundamental theoretical sense, we do not really understand. While we can thoroughly analyze some quantum-optical aspects of CIM component device behavior in the small size regime,[27][28][29] we lack a crisp understanding of how the physical dynamics of large CIMs relate to the computational complexity of combinatorial optimization. Promising experimental benchmarking results[30] are thus driving theoretical studies aimed at better elucidating fundamental operating principles of the CIM architecture and at enabling confident predictions of future scaling potential. We thus face complementary obstacles to those of mainstream quantum computing, in which we have long had theoretical analyses pointing to exponential speedups while even small-scale implementations have required sustained laboratory efforts over several decades.

What is the effective search mechanism of large-scale CIM? Are quantum effects decisive for the performance of current and near-term MFB-CIM prototypes, and if not, could existing architectures and algorithms be generalized to realize quantum performance enhancements? Can we relate exponential gain (as understood from a quantum optics perspective) to features of the phase portraits of CIMs viewed as dynamical systems, and thereby rationalize its role in facilitating rapid evolution towards states with low Ising energy? Can we rationally design better strategies for varying the pump strength?

Generally speaking, CIM may be viewed as an approach to mapping combinatorial (discrete variable) optimization problems into physical dynamics on a continuous variable space, in which the dynamics can furthermore be modulated to evolve/bifurcate the phase portrait during an individual optimization trajectory. The overarching problem of CIM algorithm design could thus be posed as choosing initial conditions for the phase-space variables together with a modulation scheme for the dynamics, such that we maximize the probability and minimize the time required to converge to states from which we can infer very good solutions to a combinatorial optimization problem instance encoded in parameters of the dynamics. While our initialization and modulation scheme obviously cannot require prior knowledge of what these very good solutions are, it should be admissible to consider strategies that depend upon inexpensive structural analyses of a given problem instance and/or real-time feedback during dynamic optimization. The structure of near-term-feasible CIM hardware places constraints on the practicable set of algorithms, while limits on our capacity to prove theorems about such complex dynamical scenarios generally restricts us to the development of heuristics rather than algorithms with performance guarantees.

We may note in passing that in addition to lifting combinatorial problems into continuous variable spaces, analog physics-based engines such as CIMs generally also embed them in larger model spaces that can be traversed in real time. The canonical CIM algorithm implicitly transitions from a linear solver to a soft-spin Ising model, and a recently-developed generalized CIM algorithm with feedback control can access a regime of fixed-amplitude Ising dynamics as well.[26] Given the central role of the optical parametric amplifier (OPA) in the CIM architecture, it stands to reason that it could be possible to transition smoothly between XY-type and Ising-type models by adjusting hardware parameters that tune the OPA between non-degenerate and degenerate operation.[31] Analog physics-based engines thus motivate a broader study of relationships among the landscapes of Ising-type optimization problems with fixed coupling coefficients but different variable types, which could further help to inform the development of generalized CIM algorithms.

The dynamics of a classical, noiseless CIM can be modeled using coupled ordinary differential equations (ODEs):

$$\frac{dx_i}{dt} = -x_i^3 + ax_i - \sum J_{ij} x_j, \tag{1}$$

where $x_i$ is the (quadrature) amplitude of the $i^{th}$ OPO mode (spin), $J_{ij}$ are the coupling coefficients defining an Ising optimization problem of interest (here we will assume $J_{ii} = 0$), and $a$ is a gain-loss parameter corresponding to the difference between the CIM's parametric (OPA) gain and its round-trip (passive linear) optical losses (e.g., [13][32]). We note that similar equations appear in the neuroscience literature for modeling neural networks (*e.g.*, [33]). In the absence of couplings among the spins ($J_{ij} \to 0$) each OPO mode independently exhibits a pitchfork bifurcation as the gain-loss parameter $a$ crosses through zero (increasing from negative to positive value), corresponding to the usual OPO "lasing" transition. With non-zero couplings however, the bifurcation set of the model is much more complicated.

In the standard CIM algorithm the $J_{ij}$ matrix is chosen to be (real) symmetric, although current hardware architectures would easily permit asymmetric implementations. With $J_{ij}$ symmetric it is possible to view the overall CIM dynamics as gradient descent in a landscape determined jointly by the individual OPO terms and the Ising potential energy. Following recent practice in related fields,[33][34] we may assess generic behavior of the above model for large problem size (large number of spins, $N$) by treating $J_{ij}$ as a random matrix whose elements are drawn i.i.d. from a zero mean Gaussian with variance $1/N$. This choice corresponds to the famous Sherrington-Kirkpatrick (SK) Ising spin glass model.[35] The origin $x_i = 0$ is clearly a fixed point of the dynamics for all parameter values, and in the loss-dominated regime ($a$ negative, and less than the smallest eigenvalue of $J_{ij}$ matrix) it is the unique stable fixed point. Assuming $J_{ij}$ is symmetric as implemented, the first bifurcation as $a$ is increased (pump power is increased) necessarily occurs as $a$ crosses the smallest eigenvalue of $J_{ij}$ and results in destabilization of the origin, with a pair of local minima emerging along positive and negative directions aligned with the eigenvector of $J_{ij}$ corresponding to this lowest eigenvalue. If we assume that the CIM is initialized at the origin (all OPO modes in vacuum) and the pump is increased gradually from zero, we may expect the spin-amplitudes to adiabatically follow this bifurcation and thus take values such that the $x_i$ are proportional to the eigenvector with a smallest eigenvalue of $J_{ij}$ just after $a$ crosses the smallest eigenvalue. The sign structure of this eigenvector is known to be a simple (although not necessarily very good) heuristic for a low-energy solution of the corresponding Ising optimization problem. For example, for the SK model, the spin configuration obtained from rounding the eigenvector with a smallest eigenvalue of $J_{ij}$ is thought to have a 16% higher energy density (energy per spin) than that of the ground state spin configuration.[36]

In the opposite regime of high pump amplitude, $a \gg |J_{ij}|$, we can infer the existence of a set of fixed points determined by the independent OPO dynamics (ignoring the $J_{ij}$ terms) with each of the $x_i$ assuming one of three possible values $\{0, \pm\sqrt{a}\}$. The leading-order effect of the coupling terms can then be considered perturbatively, leading to the conclusion[37] that the subset of fixed points without any zero values among the $x_i$ are local minima having energies

$$E(\overline{x}) = -\frac{a^2}{4} + \sum_{i,j} J_{ij} \frac{\overline{x}_i}{|\overline{x}_i|} \frac{\overline{x}_j}{|\overline{x}_j|} + O(a^{-3}), \tag{2}$$

with energy-distance relation

$$E(\bar{x}) = -\frac{1}{4} \sum_i \bar{x}_i^4.$$

Here the bar above $x$ means an ensemble average over many trajectories, when there exists stochastic noise in the system.

It follows that the global minimum spin configuration for the Ising problem instance encoded by $J_{ij}$ can be inferred from the sign structure of the local minimum lying at greatest distance from the origin, and that very good solutions can similarly be inferred from local minima at large squared-radius. We may see in this some validation of the foundational physical intuition that in a network of OPOs coupled according to a set of $J_{ij}$ coefficients, the "strongest" (largest amplitude, for a given pump strength) collective mode of oscillation should correspond somehow with an optimum solution (having minimum value of the $J_{ij}$ coupling term) of an Ising problem defined by these $J_{ij}$.

A big picture thus emerges in which initialization at the origin (all OPOs in vacuum) and adiabatic increase of the pump amplitude induces a transition between a low-pump regime in which the spin-amplitudes assume a sign structure determined by the minimum eigenvector of $J_{ij}$, and a high-pump regime in which good Ising solutions are encoded in the sign structures of minima sitting at greatest distance from the origin[37]. Apparently, complex things happen in the intermediate regime. Qualitatively speaking, the gradual increase of $a$ in the above equations of motion induces a sequence of bifurcations that modify the phase portrait in which the CIM state evolves. In simple cases, the state variables $x_i$ could follow an "adiabatic trajectory" that connects the origin (at zero pump amplitude) to a fixed point in the high-pump regime (asymptotic in large $a$) whose sign structure yields a heuristic solution to the Ising optimization. In general, one observes that such adiabatic trajectories include sign flips relative to the first-bifurcated state proportional to the eigenvector with a smallest eigenvalue of $J_{ij}$. In a non-negligible fraction of cases, as revealed by numerical characterization of the bifurcation set for randomly-generated $J_{ij}$ with $N \sim 10^2$, the adiabatic trajectory starting from the origin is at some point interrupted by a subcritical bifurcation that destabilizes the local minimum being followed without creating other local minima in the immediate neighborhood. (Indeed, some period of evolution along an unstable manifold would seem to be required for the observation of a lasing transition with exponential gain.) For such problem instances, a fiduciary evolution of the CIM state cannot be directly inferred from computation of fixed-point trajectories as a function of $a$.

Generally speaking, in the "near-threshold" regime with $a \sim 0$ we may expect the CIM to exhibit "glassy" dynamics with pervasive marginally-stable local minima, and as a consequence the actual solution trajectory followed in a real experimental run could depend strongly on exogenous factors such as technical noise and instabilities. Hence it is not clear whether we should expect the type of adiabatic trajectory described above to occur commonly, in practice. Indeed, fluctuations could potentially induce accidental asymmetries in the implementation of the $J_{ij}$ coupling term, which could in turn induce chaotic transients that significantly affect the optimization dynamics. We note that the existence of a chaotic phase has been predicted[33] on the basis of mean-field theory (in the sense of statistical mechanics) for a model similar to the CIM model considered here, but with a fully random coupling matrix without symmetry constraint. Characterization of the phase diagram for near-symmetric $J_{ij}$ (nominally symmetric but with small asymmetric perturbations) seems feasible and is currently being studied.[38] It is tempting to ask whether a glassy phase portrait for the classical ODE model in the near-threshold regime could correspond in some way with non-classical behavior observed in full quantum simulations of optical delay line coupled coherent Ising machine (ODL-CIM) models near threshold, as reviewed in the next section. It seems natural to conjecture that quantum uncertainties associated with anti-squeezing below threshold could induce coherent spreading over a glassy landscape with numerous marginal minima, with associated buildup of quantum correlation among spin-amplitudes.

The above picture calls attention to a need to understand the topological nature of the phase portrait and its evolution as the pump amplitude, $a$, is varied. Indeed, we may restate in some sense the abstract formulation of the CIM algorithm design problem: Can we find a strategy for modulating the CIM dynamics in a way that enables us to predict (without prior knowledge of actual solutions) how to initialize the spin-amplitudes such that they are guided into the basin of attraction of the largest-radius minimum in the high pump regime? Or into one of the basins of attraction of a class of acceptably large-radius minima (corresponding to very good solutions)? Of course, an additional auxiliary design goal will be to guide the CIM state evolution in such a way that the asymptotic sign structure is reached quickly.

In the near/below-threshold regime, we may anticipate at least two general features of the phase portrait that could present obstacles to rapid equilibration. One would be the afore-mentioned prevalence of marginal local minima (having eigenvalues with very small or vanishing real part), but another would be a prevalence of low-index saddle points. Trajectories within either type of phase portrait could display intermittent dynamics that impede gradient-descent towards states of lower energy. Focusing on the below-threshold regime in which the Ising-interaction energy term may still dominate the phase portrait topology, we might infer from works such as [39] that for large $N$ with $J_{ij}$ symmetric-random-Gaussian, fixed points lying well above the minimum energy could dominantly be saddles with strong correlation between the energy of a fixed point and its index (fraction of unstable eigenvalues), as discussed in [34]. If such features of the landscape for random-Gaussian $J_{ij}$ generalize to instances of practical interest, as a gradient-descent trajectory approaches phase space regions of lower and lower energy, results from [34][39] suggest that the rate of descent could become limited by escape times from low-index saddles whose eigenvalues are not necessarily small, but whose local unstable manifold may have dimension small relative to $N$.

One wonders whether there might be CIM dynamical regimes in which the gradient-descent trajectory takes on the character of an "instanton cascade" that visits (neighborhoods of) a sequence of saddle points with decreasing index,[40] leading finally to a local minimum at low energy. If such dynamics actually occurs in relevant operating regimes for CIM, we may speculate as to whether the overall gradient descent process including stochastic driving terms (caused by classical-technical or quantum noise) could reasonably be abstracted as probability (or quantum probability-amplitude) flow on a graph. Here the nodes of the graph would represent fixed points and the edges would represent heteroclinic orbits, with the precise structure of the graph of course determined by $a$ and $J_{ij}$. If the graph for a given problem-instance exhibits loops, we could ask whether interference effects might lead to different transport rates for quantum versus classical flows (as in quantum random walks[41]). Such effects, if they exist, would provide intriguing examples of ways in which limited transient entanglement (localized to the phase space-time volume of traversing a graph loop) could impact optimization dynamics in a computational architecture.

Quantum noise correlation for parallel search

The first CIM demonstrated in 2014 uses $(N-1)$ optical delay lines to all-to-all couple N-OPO pulses circulating in a ring cavity according to the target Hamiltonian $\mathcal{H} = \sum J_{ij} x_i x_j$, where $x_i$ is the in-phase amplitude of $i$-th OPO pulse (see Fig. 1 in [14]). A coupling field $I_i$ is chosen as the gradient of the potential, $I_i \propto -\partial \mathcal{H}/\partial x_i$, where the analog (quadrature) amplitude $x_i$ represents the binary spin variable by $x_i = \sigma_i |x_i|$. When an Ising coupling coefficient $J_{ij}$ between the $i$-th and $j$-th OPO pulses is positive (ferromagnetic coupling), an optical delay line realizes in-phase coupling between (internal) $i$-th and (externally injected) $j$-th OPO pulse amplitudes (see Fig. 1(a)). The OPO pulses incident upon an extraction beam splitter (XBS) carry anti-squeezed vacuum noise along a generalized coordinate $x$ at below threshold, which produces a positive noise correlation between transmitted and reflected OPO pulses after XBS, while vacuum noise from the XBS open port is negligible compared to the anti-squeezed noise of incident OPO pulse. Positive noise correlation is similarly established between the $i$-th and $j$-th OPO pulses after combining the injected $j$-th OPO pulse and internal $i$-th OPO pulse at an injection beam splitter (IBS), as shown in Fig. 1(a). When the Ising coupling coefficient $J_{ij}$ is negative (anti-ferromagnetic coupling), the optical delay line realizes an out-of-phase coupling between the $i$-th and $j$-th OPO pulse amplitudes. This setup of the optical delay line results in the negative noise correlation between the two OPO pulse amplitudes.

Below threshold, each OPO pulse is in an anti-squeezed vacuum state which can be interpreted as a linear superposition (not statistical mixture) of generalized coordinate eigenstates, $\sum_n c_n |x_n\rangle$, if the decoherence effect by linear cavity loss is neglected. In fact, quantum coherence between different $|x\rangle$ eigenstates is very robust against small linear loss.[23] Figure 1(b) shows the quantum noise trajectory in $\langle \Delta \widehat{X}^2 \rangle$ and $\langle \Delta \widehat{P}^2 \rangle$ phase space. The uncertainty product stays close to the Heisenberg limit, with a very small excess factor of less than 30%, during an entire computation process, which suggests the purity of an OPO state is well maintained.[42] Therefore, the above mentioned positive/negative noise correlation between two OPO pulses depending on ferromagnetic/anti-ferromagnetic coupling, implements a sort of quantum parallel search. That is, if the two OPO pulses couple ferromagnetically, the formed positive quantum noise correlation prefers ferromagnetic phase states $|0\rangle_i |0\rangle_j$ and $|\pi\rangle_i |\pi\rangle_j$, where $|0\rangle = \int_0^\infty c_X |x\rangle \mathbf{dx}$ and $|\pi\rangle = \int_{-\infty}^0 c_X |x\rangle \mathbf{dx}$. If two OPO pulses couple anti-ferromagnetically, the formed negative quantum noise correlation prefers anti-ferromagnetic phase states $|0\rangle_i |\pi\rangle_j$ and $|\pi\rangle_i |0\rangle_j$.

Entanglement and quantum discord between two OPO pulses can be computed to demonstrate such quantum noise correlations.[27][28][29] Figure 1(c) and (d) show the degrees of entanglement and quantum discord versus normalized pump rate $p = \mathcal{E}/\mathcal{E}_{th}$, whew $\mathcal{E}$ is a pump field amplitude incident upon the nonlinear crystal and $\mathcal{E}_{th}$ is that at the oscillation threshold for a solitary OPO, for an ODL-CIM with $N = 2$ pulses.[29] In Fig. 1(c), it is shown that Duan-Giedke-Cirac-Zoller entanglement criterion[43] is satisfied at all pump rates. In Fig 1(d), it is shown that Adesso-Datta quantum discord criterion[44] is also satisfied at all pump rate.[29] Both results on entanglement and quantum discord demonstrate maximal quantum noise correlation formed at threshold pump rate $p = 1$. On the other hand, if a (fictitious) mean-field without quantum noise is assumed to couple two OPO pulses, there exists no quantum correlation below or above threshold, as shown by open circles in Fig. 1(d).

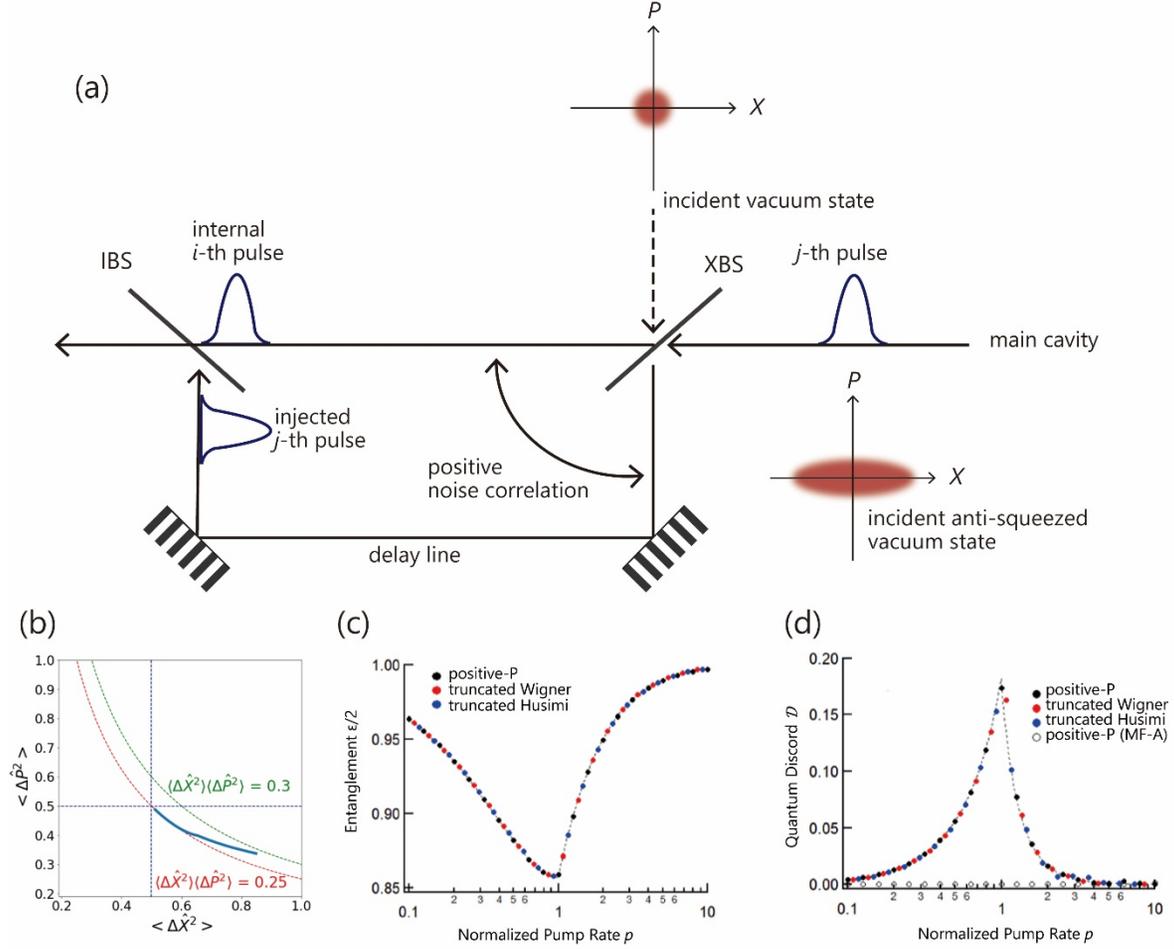

**FIG. 1.** (a) An optical delay line couples two OPO pulses in ODL-CIM.[14] (b) Variances $\langle\Delta\hat{X}^2\rangle$ and $\langle\Delta\hat{P}^2\rangle$ in a MFB-CIM with $N = 16$ OPO pulses. The uncertainty product deviates from the Heisenberg limit by less than 30%.[42] (c) Duan-Giedke-Cirac-Zoller inseparability criterion ($\varepsilon/2 < 1$) vs. normalized pump rate $p = \mathcal{E}/\mathcal{E}_{th}$. Numerical simulations are performed by the positive-$P$, truncated-Wigner and truncated-Husimi stochastic differential equations (SDE). The dashed line represents an analytical solution.[29] (d) Adesso-Datta quantum discord criterion ($D > 0$) vs. normalized pump rate $p$. The above three SDEs and the analytical result predict the identical quantum discord, while the mean-field coupling approximation (MF-A) predicts no quantum discord.[29]

Note that vacuum noise incident from an open port of XBS (See Fig. 1(a)) creates an opposite noise correlation between the internal and external OPO pulses, so that it always degrades the preferred quantum noise correlation among the two OPO pulses after IBS. Thus, squeezing the vacuum noise at open port of XBS is expected to improve the quantum search performance of an ODL-CIM, which is indeed confirmed in the numerical simulation.[28]

The second generation of CIM demonstrated in 2016 employs a measurement-feedback circuit to all-to-all couple the N-OPO pulses (see Fig. 1 of [16]). The (quadrature) amplitude $x_j$ of a reflected OPO pulse $j$ after XBS is measured by an optical homodyne detector and the measurement result (inferred amplitude) $\tilde{x}_j$ is multiplied against the Ising coupling coefficient $J_{ij}$ and summed over all $j$ pulses in electronic digital circuitry, which produces an overall feedback signal $\sum_j J_{ij} \tilde{x}_j$ for the $i$-th internal OPO pulse. This analog electrical signal is imposed on the amplitude of a coherent optical feedback signal, which is injected into the target OPO pulse $i$ by IBS. In this MFB-CIM operating below threshold, if a homodyne measurement result $\tilde{x}_j$ is positive and incident vacuum noise from the open port of XBS is negligible, the average amplitude of the internal OPO pulse $j$ is shifted (jumped) to a positive direction by the projection property of such an indirect quantum measurement[45], as shown in Fig. 2. Depending on the value of a feedback signal $J_{ij} \tilde{x}_j$, we can introduce either positive or negative displacement for the center position of the target OPO pulse $i$. In this way, depending on the sign of $J_{ij}$, we can implement either positive correlation or negative correlation between the two average amplitudes $\langle x_i \rangle$ and $\langle x_j \rangle$ for ferromagnetic or anti-ferromagnetic coupling, respectively. Note that a MFB-CIM does not produce entanglement among OPO pulses but generates quantum discord if the density operator is defined as an ensemble over many measurement records.[46] A normalized correlation function $N = \langle \Delta \hat{X}_1 \Delta \hat{X}_2 \rangle / \sqrt{\langle \Delta \hat{X}_1^2 \rangle \langle \Delta \hat{X}_2^2 \rangle}$ is an appropriate metric for quantifying such measurement-feedback induced search performance, the degree of which is shown to govern final success probability of MFB-CIM more directly than the quantum discord. In general, a MFB-CIM has a larger normalized correlation function and higher success probability than an ODL-CIM.[46]

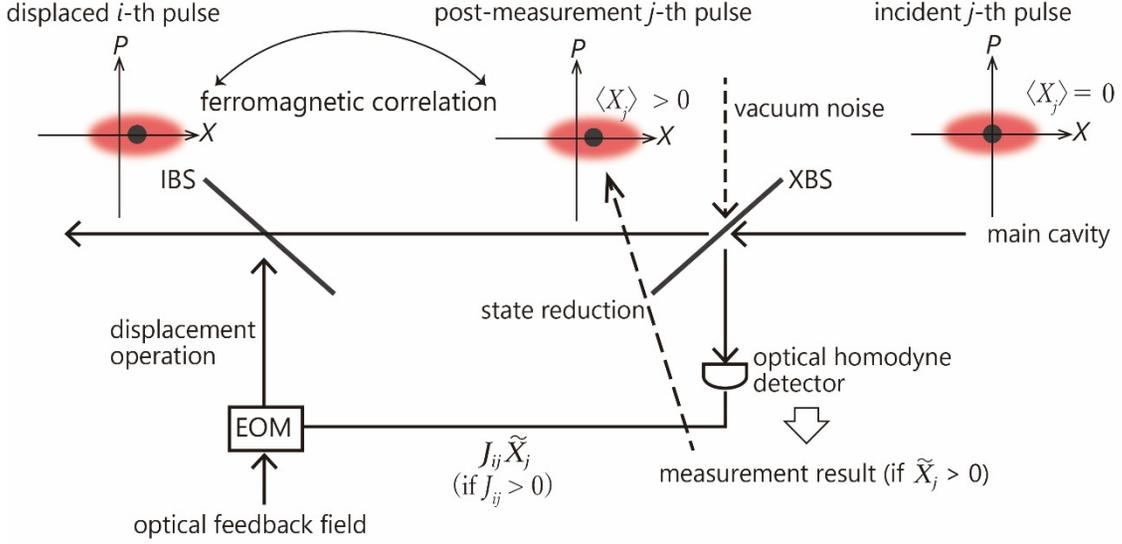

**FIG. 2.** Formation of a ferromagnetic correlation between two OPO pulses in MFB-CIM.[15][16] This example illustrates the noise distributions of the two OPO pulses when the Ising coupling is ferromagnetic ($J_{ij} > 0$) and the measurement result for the $j$-th pulse is $\widetilde{X}_j > 0$.

In both ODL-CIM and MFB-CIM, anti-squeezed noise below threshold makes it possible to search for a lowest-loss ground state as well as low-loss excited states before the OPO network reaches threshold. The numerical simulation result shown in Fig. 3 demonstrates the three step computation of CIM.[28] We study a $N = 16$ one-dimensional lattice with a nearest-neighbor anti-ferromagnetic coupling and periodic boundary condition ($x_1 = x_{17}$), for which the two degenerate ground states are $|0\rangle_1|\pi\rangle_2 \cdots\cdots |0\rangle_{15}|\pi\rangle_{16}$ and $|\pi\rangle_1|0\rangle_2 \cdots\cdots |\pi\rangle_{15}|0\rangle_{16}$. Before a pump field $\mathcal{E}$ is switched on ($t \leq 0$ in Fig. 3), all OPOs are in vacuum states, for which optical homodyne measurement results provide a simple random guess for each spin and the corresponding success rate is $P_s = 1/2^{16} \sim 10^{-5}$ as shown by the horizontal solid line in Fig. 3. We assume that vacuum noise incident from the open port of XBS is squeezed by 10 dB in ODL-CIM. When the external pump rate is linearly increased from below to above threshold, the probability of finding the two degenerate ground states is increased by two orders of magnitude above the initial success probability. This enhanced success probability stems from the formation of quantum noise correlation among 16 OPO pulses at below threshold. The probability of finding high-loss excited states, which are not shown in Fig. 3, is deceased to below the initial value. This "quantum preparation" is rewarded at the threshold bifurcation point. When the pump rate reaches threshold, one of the ground states ($|0\rangle_1|\pi\rangle_2 \cdots\cdots |\pi\rangle_{16}$) in the case of Fig. 3 is selected as a single oscillation mode, while the other ground state ($|\pi\rangle_1|0\rangle_2 \cdots\cdots |0\rangle_{16}$) as well as all excited states are not selected. This is not a standard single oscillator bifurcation but a collective phenomenon among $N = 16$ OPO pulses due to the existence of anti-ferromagnetic noise correlation. Above threshold, the probability of finding the selected ground state is exponentially increased, while those of finding the unselected ground state as well as all excited states are exponentially suppressed in a time scale of the order of signal photon lifetime. Such exponential amplification and attenuation of the probabilities is a unique advantage of a gain-dissipative computing machine, which is absent in a standard (unitary gate based) quantum computing system. For example, the Grover search algorithm utilizes a unitary rotation of state vectors and can amplify the target state amplitude only linearly.[25] However, this difference does not mean the computational time of CIM is sub-exponential for hard instances. Recent numerical studies suggest that CIM has an improved but still exponentially scaling time.[30] Note that if we stop increasing the pump rate just above threshold, the probability of finding either one of the ground states is less than 1%. Pitchfork bifurcation followed by exponential amplitude amplification plays a crucial role in realizing high success probability in a short time.

For hard instances of combinatorial optimization problems, in which excited states form numerous local minima, the above quantum search alone is not sufficient to guarantee a high success probability.[30] In the next section, another CIM with error correction feedback is introduced to cope with such hard instances.[26] An alternative approach has been recently proposed.[42] If a pump rate is held just below threshold (corresponding to $t \sim 60$ in Fig. 3), the lowest-loss ground states and low-loss excited states (fine solutions) have enhanced probabilities while high-loss excited states have suppressed probabilities. By using a MFB-CIM, the optimum as well as good sub-optimal solutions are selectively sampled through an indirect measurement in each round trip of the OPO pulses. This latter approach is particularly attractive if the computational goal is to sample not only optimum solutions but also semi-optimum solutions.

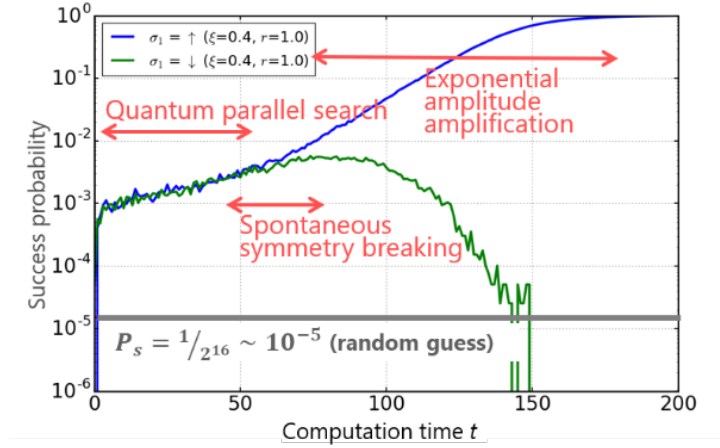

**FIG. 3.** The probabilities of finding two degenerate ground states in ODL-CIM for a one-dimension lattice with nearest-neighbor anti-ferromagnetic coupling and periodic boundary condition ($x_1 = x_{17}$).[28] Many trajectories produced by the numerical simulation with truncated-Wigner SDE are post-selected by the final state $|0\rangle_1 |\pi\rangle_2 \cdots\cdots |0\rangle_{15} |\pi\rangle_{16}$ and ensemble averaged.

Destabilization of local minima

The measurement-feedback coherent Ising machine has been previously described as a quantum analog device that finishes computation in a classical digital device, in which the amplitude of a selected low energy spin configuration is exponentially amplified.[22][23] During computation, the sign of the measured in-phase component, noted $\tilde{x}_i$ with $\tilde{x}_i \in \mathbb{R}$, is associated with the boolean variable $\sigma_i$ of an Ising problem (whereas the quadrature-phase component decays to zero). A detailed model of the system's dynamics is given by the master equation of the density operator $\rho$ that is conditioned on measurement results[47][48] which describes the processes of parametric amplification (exchange of one pump photon into two signal photons), saturation (signal photons are converted back into pump photons), wavepacket reduction due to measurement, and feedback injection that is used for implementing the Ising coupling. For the sake of computational tractability, truncated Wigner[28] or the positive-P representation[49] can be used with Itoh calculus for approximating the quantum state to a probability distribution $P(x_i)$ with $P(x_i) \approx Tr[|x_i\rangle\langle x_i|\rho]$ from which the measured in-phase component $\tilde{x}_i$ can be calculated with $\tilde{x}_i = \langle x_i \rangle + \gamma \eta_i$ where $\langle x_i \rangle = \int x_i P(x_i) dx_i$ and $\eta_i$ are uncorrelated increments with amplitude $\gamma > 0$.

Although gain saturation and dissipation can, in principle, induce squeezing and non-Gaussian states[50] that would justify describing the time-evolution of the higher moments of the probability distribution P, it is insightful to limit our description to its first moment (the average $\langle x_i \rangle$) in order to explain computation achieved by the machine in the classical regime. This approximation is justified when the state of each OPO remains sufficiently close to a coherent state during the whole computation process. In this case, the effect of gain saturation and dissipation on the average $\langle x_i \rangle$ can be modeled as a non-linear function $x \mapsto f(x)$ and the feedback injection is given as $\beta_i \Sigma_j J_{ij} g(\langle x_i \rangle + \gamma \eta_j)$ where $f$ and $g$ are sigmoid functions, $J_{ij}$ the Ising couplings, and $\beta_i$ represents the amplitude of the coupling. When the amplitudes $|\langle x_i \rangle|$ of OPO signals are much larger that the noise amplitude $\gamma$, the system can be described by simple differential equations given as $(d/dt)\langle x_i \rangle = f(\langle x_i \rangle) + \beta \Sigma_j j_{ij} g(\langle x_j \rangle)$ for which we set $\beta_i = \beta$, $\forall i$, and it can be shown that the time-evolution of the system is a motion in state space that seeks out minima of a potential function (or Lyapunov function) $V$ given as $V = V_b(y) + \beta H(y)$ where $V_b$ is a bistable potential with $V_b(y) = -\Sigma_i \int^{y_i} f(g^{-1}(y))dy$ and $H(y) = -(1/2)\Sigma_{ij} j_{ij} y_i y_j$ is the extension of the Ising Hamiltonian in the real space with $y_i = g(\langle x_i \rangle)$.[21][51] The connection between such nonlinear differential equations and the Ising Hamiltonian has been used in various models such as in the "soft" spin description of frustrated spin systems[52] or the Hopfield-Tank neural networks[51] for solving NP-hard combinatorial optimization problems. Moreover, an analogy with the mean-field theory of spin glasses can be made by recognizing that the steady-states of these nonlinear equations correspond to the solution of the "naive" Thouless-Anderson-Palmer (TAP) equations[53] which arise from the mean-field description of Sherrington-Kirkpatrick spin glasses in the limit of large number of spins $N$ and are given as $\langle \sigma_i \rangle = \tanh((1/T)\Sigma_j J_{ij} \langle \sigma_j \rangle)$ with $\langle \sigma_i \rangle$ the thermal average at temperature $T$ of the Ising spin (by setting $f(x) = a\tanh(x)$ and $g(x) = x$). This analogy suggests that the parameter $\beta$ can be interpreted as inverse temperature in the thermodynamic limit when the Onsager reaction term is discarded.[53] At $\beta = 0$ ($T \to \infty$), the only stable state of the CIM is $\langle x_i \rangle = 0$, for which any spin configuration is equiprobable, whereas at $\beta \to \infty$ ($T = 0$), the state remains trapped for an infinite time in local minima. We will discuss in much more detail analogies between CIM dynamics and TAP equations, and also belief and survey propagation, in the special case of the SK model in the next section.

In the case of spin glasses, statistical analysis of TAP equations suggests that the free energy landscape has an exponentially large number of solutions near zero temperature[54] and we can expect similar statistics for the potential $V$ when $\beta \to \infty$. In order to reduce the probability of the CIM to get trapped in one of the local minima of $V$, it has been proposed to gradually increase $\beta$, the coupling strength, during computation.[16] This heuristic, that we call open-loop CIM in the following, is similar to mean-field annealing[55] and consists in letting the system seeks out minima of a potential function $V$ that is gradually transformed from monostable to multi-stable (see Fig. 4(a) and (b1)). Contrarily to the quantum adiabatic theorem[56] or the convergence theorem of simulated annealing,[57] there is however no guarantee that a sufficiently slow deformation of $V$ will ensure convergence to the configuration of lowest Ising Hamiltonian. In fact, linear stability analysis suggests on the contrary that the first state other than vacuum state ($\langle x_i \rangle = 0$, $\forall\, i$) to become stable as $\beta$ is increased does not correspond to the ground-state. Moreover, added noise $\eta_i$ may not be sufficient for ensuring convergence:[58] it is possible to seek for global convergence to the minima of the potential $V$ by reducing gradually the amplitude of the noise $\gamma$ (with $\gamma(t)^2 \sim c/\log(2+t)$ and $c$ real constant sufficiently large,[59] but the global minima of the potential $V(y)$ do not generally correspond to that of the Ising Hamiltonians $H(\boldsymbol{\sigma})$ at a fixed $\beta$.[13][21] This discrepancy between the minima of the potential $V$ and Ising Hamiltonian $H$ can be understood by noting that the field amplitudes $\langle x_i \rangle$ are not all equal (or homogeneous) at the steady-state, that is $\langle x_i \rangle = \sigma_i \sqrt{a} + \delta_i$ where $\delta_i$ is the variation of the $i$-th OPO amplitude with $\delta_i \neq \delta_j$ and $\sqrt{a}$ a reference amplitude defined such that $\Sigma_j \delta_j = 0$. Because of the heterogeneity in amplitude, the minima of $V(\langle x \rangle) = V(\boldsymbol{\sigma}\sqrt{a} + \boldsymbol{\delta})$ do not correspond to that of $H(\boldsymbol{\sigma})$ in general. Consequently, it is necessary in practice to run the open-loop CIM from multiple initial conditions in order to find the ground-state configuration.

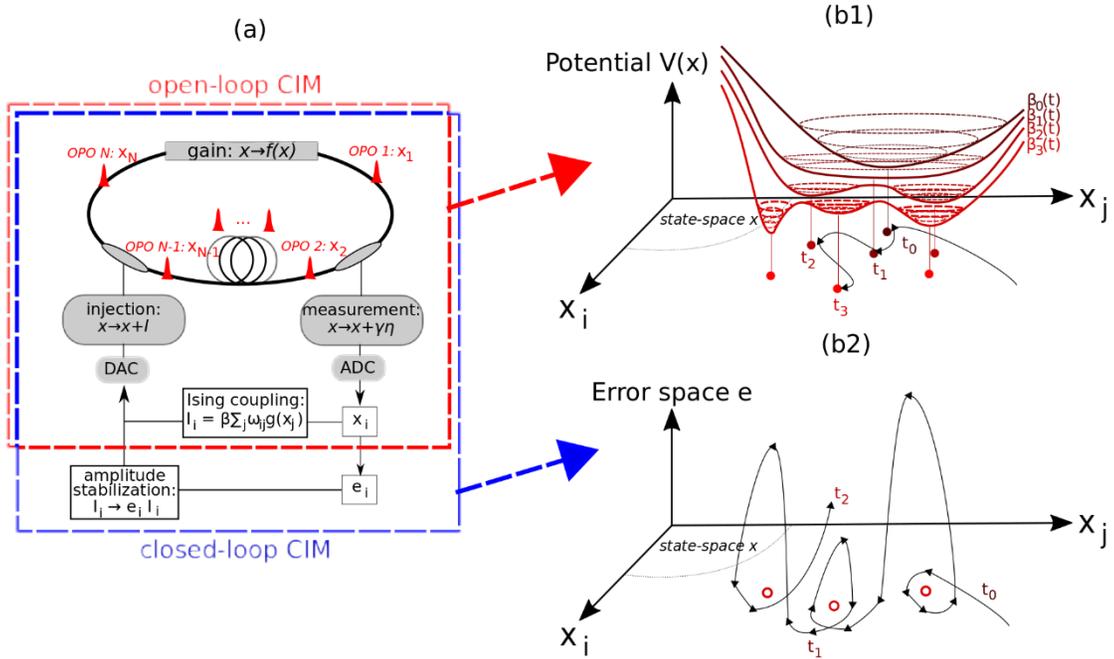

**FIG. 4.** (a) Schematic representation of the open- and closed-loop measurement-feedback coherent Ising machine which computational principle in the mean-field approximation are based on two different types of dynamical systems: gradual deformation of a potential function $V$ (b1) and chaotic-like dynamics (b2), respectively.

Because the benefits of using an analog state for finding the ground-state spin configurations of the Ising Hamiltonian is offset by the negative impact of its improper mapping to the potential function $V$, we have proposed to utilize supplementary dynamics that are not related to the gradient descent of a potential function but ensure that the global minima of $H$ are reached rapidly. In Ref [26], an error correction feedback loop has been proposed whose role is to reduce the amplitude heterogeneity $\delta_i$ by forcing squared amplitudes $\langle x_i \rangle^2$ to become all equal to a target value $a$, thus forcing the measurement-feedback coupling $\{\Sigma_j J_{ij} g(\langle x_j \rangle)\}_i$ to be colinear with the Ising internal field $\boldsymbol{h}$ with $h_i = \Sigma_j J_{ij} \sigma_j$. This can notably be achieved by introducing error signals, noted $e_i$ with $e_i \in \mathbb{R}$, that modulate the coupling strength $\beta_i$ (or "effective" inverse temperature) of the $i$-th OPO such that $\beta_i = \beta e_i(t)$ and the time-evolution of $e_i$ given as $(d/dt)e_i = -\xi(g(\langle x_i \rangle)^2 - a)e_i$ where $\xi$ is the rate of change of error variables with respect to the signal field. This mode of operation is called closed-loop CIM and can be realized experimentally by simulating the dynamics of the error variables $e_i$ using the FPGA used in the measurement-feedback CIM for calculation of the Ising coupling[16] (see Fig. 4(a)). Note that the concept of amplitude heterogeneity error correction has also been recently proposed in [20] and extended to other systems such as the XY model.[20][60][61]

In the case of the closed-loop CIM, the system exhibits steady-states only at the local minima of $H$.[26] The stability of each local minima can be controlled by setting the target amplitude a as follows: the dimension of the unstable manifold $N_u$ (where $N_u$ is the number of unstable directions) at fixed points corresponding to local minima $\boldsymbol{\sigma}$ of the Ising Hamiltonian is equal to the number of eigenvalues $\mu_i(\boldsymbol{\sigma})$ that are such that $\mu_i(\boldsymbol{\sigma}) > F(a)$ where $\mu_i(\boldsymbol{\sigma})$ are the eigenvalues of the matrix $\{J_{ij}/|h_i|\}_i$ (with internal field $h_i = \Sigma_j J_{ij} \sigma_j$) and $F$ a function shown in Fig. 5(a). The parameter $a$ can be set such that all local minima (including the ground-state) are unstable such that the dynamics cannot become trapped in any fixed point attractors. The system then exhibits chaotic dynamics that explores successively local minima. Note that the use of chaotic dynamics for solving Ising problems has been discussed previously,[24][62] notably in the context of neural networks, and it has been argued that chaotic fluctuations may possess better properties than Brownian noise for escaping from local minima traps. In the case of the closed-loop CIM, the chaotic dynamics is not merely used as a replacement to noise. Rather, the interaction between nonlinear gain saturation and error-correction allows a greater reduction of the unstable manifold dimension of states associated with lower Ising Hamiltonian (see Fig. 5(b)). Comparison between Fig. 5(c1,d1,e1) and (c2,d2,e2) indeed shows that the dynamics of closed-loop CIM samples more efficiently from lower-energy states when the gain saturation is nonlinear compared to the case without nonlinear saturation, respectively.

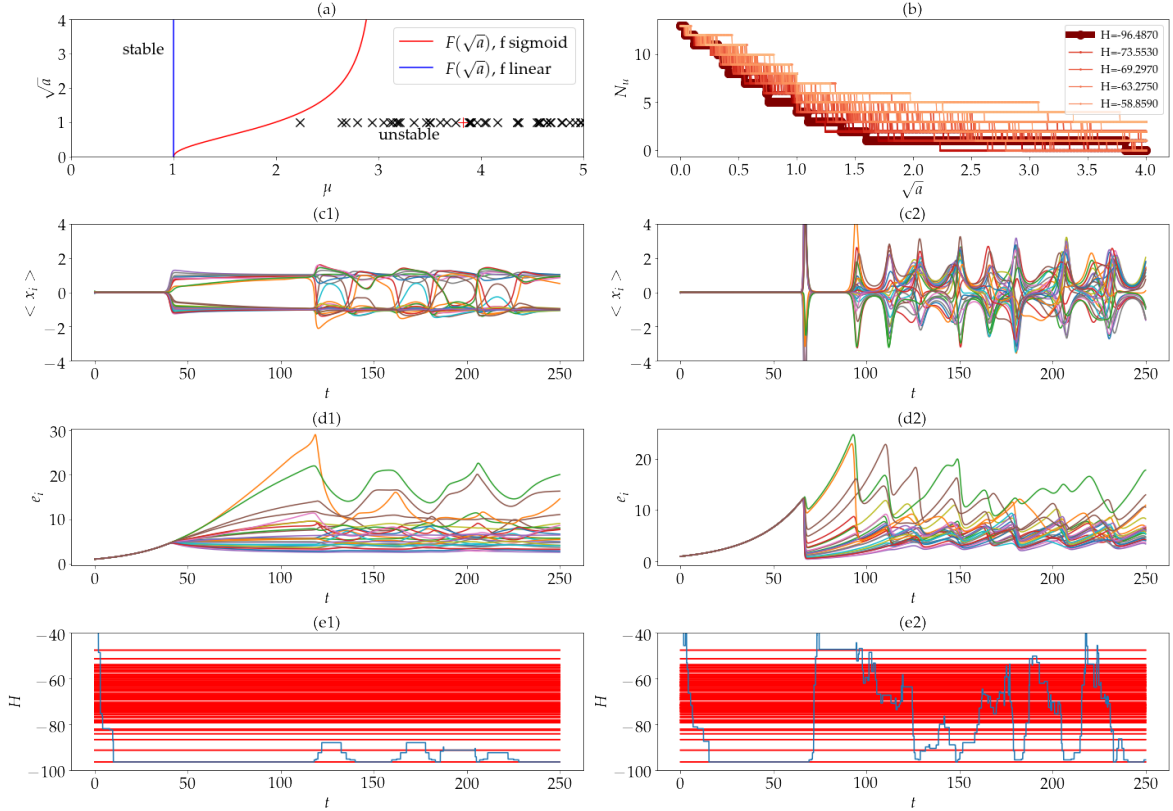

**FIG. 5.** (a) Stability of local minima in the space $\{\sqrt{a}, \mu\}$ in the case of f sigmoid and linear, i.e., with and without gain saturation, respectively. The red and black crosses correspond to the maximum values of $\mu_i(\sigma)$ for the ground-state and excited states, respectively, of an example spin-glass instance with $N = 26$ spins. (b) Dimension of the unstable manifold vs. the target amplitude $\sqrt{a}$ for the various local minima with Ising Hamiltonian $H$ shown by the color gradient in the case of $f$ sigmoid. (c1), (d1), and (e1) show the time-evolution of in-phase components $\langle x_i \rangle$, error variables $e_i$, and Ising Hamiltonian in the case of f sigmoid. (c2,d2,e2) same as (c1,d1,e1) in the case of f linear. In (e1,e2), the red lines show the Ising Hamiltonian of the local minima.

Generally, the asymmetric coupling between in-phase components and error signals possibly results in the creation of limit cycles or chaotic attractors that can trap the dynamics in a region that does not include the global minima of the Ising Hamiltonian. A possible approach to prevent the system from getting trapped in such non-trivial attractors is to dynamically modulate the target amplitude such that the rate of divergence of the velocity vector field remains positive.[26] This implies that volumes along the flow never contract which, in turn, prevents the existence of any attractor.

Figure 6(a) shows that the closed-loop CIM can find the ground-states of Sherrington-

Kirkpatrick spin glass problems with high success probability using a single run even for larger system size. Moreover, the correction of amplitude heterogeneity allows for a significant decrease in the time-to-solution compared to the open-loop case which is evaluated by calculating the number of cavity round-trip of the OPO pulses, called number of iterations and noted $n_s$, to find the ground-state configurations with 99% success probability (see Fig. 6(b)). Because there is no theoretical guarantee that the system will find configuration with Ising Hamiltonian at a ratio of the ground-state after a given computational time and the closed-loop CIM is thus classified as a heuristic method. In order to compare it with other state-of-the-art heuristics, the proposed scheme has been applied to solving instances of standard benchmarks (such as the G-set) by comparing time-to-solutions for reaching a predefined target such as the ground-state energy, if it is known, or the smallest energy known (i.e., published), otherwise. The amplitude heterogeneity error correction scheme can in particular find lower energy configurations of MAXCUT problems from the G-set of similar quality as the state-of-the-art solver, called BLS[63] (see the supplementary material of ref [26] for details). Moreover, the averaged time-to-solution obtained using the proposed scheme are similar to the ones obtained using BLS when simulated on a desktop computer, but are expected to be 100-1000 times smaller in the case of an implementation on the coherent Ising machine.

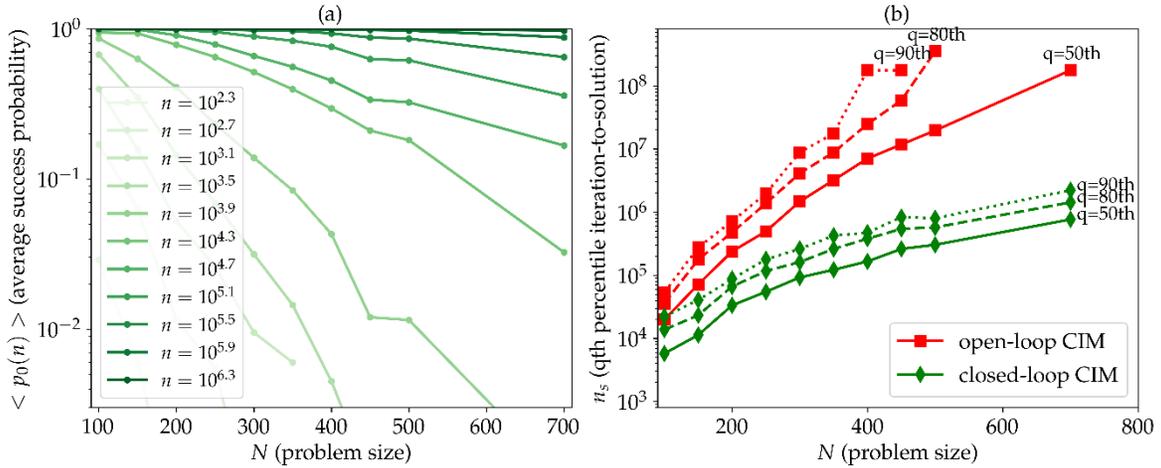

**FIG. 6.** (a) Success probability $\langle p_0(n) \rangle$ of finding the ground-state configuration after $n$ iterations of a single run averaged over 100 randomly generated Sherrington-Kirkpatrick spin glass instances and (b) 50th, 80th, and 90th percentiles of the iteration-to-solution distribution where n$_s$ of a given instance is given as $n_s = \min_n \{n_s(n)\}$ and $n_s(n) = n \log(1 - 0.99) / \log(1 - p_0(n))$ for the open-loop (in red) and closed-loop CIM (in green).

Qualitative parallels between the CIM, belief propagation and survey propagation

As we have noted above, the CIM approach to solving combinatorial optimization problems over binary valued spin variables $\sigma_i = \pm 1$ can be understood in terms of two key steps. First, in the classical limit of the CIM, the binary valued spin variables $\sigma_i$ are promoted to analog variables $x_i$ reflecting the (quadrature) amplitude of the $i^{th}$ OPO mode and the classical CIM dynamics over the variables $x_i$ can be described by a nonlinear differential equation (Eq. 1). Second, in a more quantum regime, the CIM implements a quantum parallel search over this space that focuses quantum amplitudes on the ground state. A qualitatively similar two step approach of state augmentation and then parallel search has also been pursued in statistics and computer science based approaches to combinatorial optimization, specifically in the forms of algorithms known as belief propagation (BP)[64] and survey propagation (SP).[65] Here we outline similarities and differences between CIM, BP and SP. Forming a bridge between these fields can help progress through the cross-pollination of ideas in two distinct ways. First, our theoretical understanding of BP and SP may provide further tools, beyond the dynamical systems theory approaches described above, to develop a theoretical understanding of CIM dynamics. Second, differences between CIM dynamics and BP and SP dynamics may provide further inspiration for the rational engineering design of modified CIM dynamics that could lead to improved performance. Indeed there is a rich literature connecting BP and SP to other ideas in statistical physics, such as the Bethe approximation, the replica method, the cavity method, and TAP equations.[66][67][68][69][70] It may also be interesting to explore connections between these ideas and the theory of CIM dynamics.

We begin by discussing BP, which is a general method for computing the marginal distribution $P(\sigma_i) = \sum_{\sigma_{/i}} P(\sigma_1, \ldots, \sigma_N)$ from a complex joint distribution over N variables $\sigma_j$ for $j = 1 \ldots N$. Here the sum over $\sigma_{/i}$ denotes a sum over all variables $\sigma_j$ for $j \neq i$. For binary spin variables $\sigma_j = \pm 1$, and for general joint distributions $P(\sigma_1, \ldots \sigma_N)$, the computation of this marginal distribution is intractable, as it involves a sum over $O(2^N)$ spin configurations. However, consider the case where the joint distribution $P(\sigma_1, \ldots \sigma_N)$ has a locally factorizable structure of the form $P(\sigma_1, \ldots \sigma_N) = \frac{1}{Z}\prod_{a=1}^{P}\psi_a(\sigma_a)$. Here we have $P$ interaction terms indexed by $a = 1, \ldots, P$, and each interaction term, or factor $\psi_a$ depends on a subset of the spins denoted by $\sigma_a$. Note that a parameter "a" represents an interaction term in this section, while it means a target intensity in the previous section. For example, in the case of Ising spin systems with pairwise interactions at inverse temperature $\beta$, each subset $a$ corresponds to a pair of spins $a = \{i, j\}$, the corresponding factor is given by $\psi_a(\sigma_a) = e^{\beta J_{ij}\sigma_i\sigma_j}$, and Z is the usual partition function that normalizes the distribution. The structure of the joint distribution in equation (3) below can be visualized as a factor graph, with $N$ circular nodes denoting the variables $\sigma_i$ and $P$ square factor nodes denoting the interactions $\psi_a$ (Fig. 7(a)). A variable node $i$ is connected to a factor node $a$ if and only if variable $i$ belongs to the subset $a$, or equivalently if the interaction term $\psi_a$ depends on $\sigma_i$.

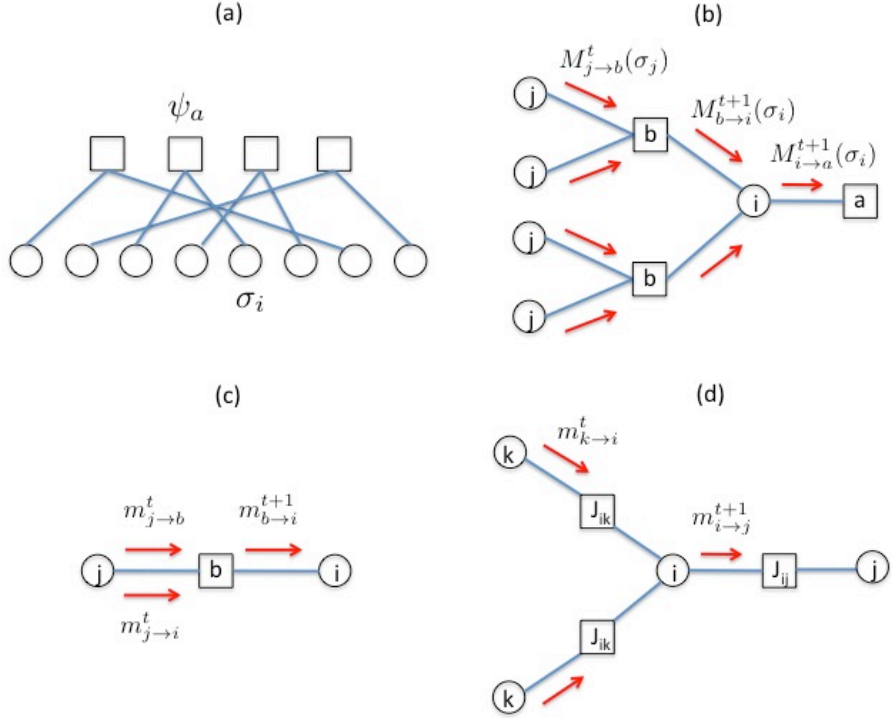

FIG 7. (a) A factor graph representation of a joint probability distribution where each square is a factor node encoding an interaction $\psi_a(\sigma_a)$. Each factor node is connected to a subset $a$ of variable nodes (circles) where each variable $\sigma_i$ is connected to factor $a$ if and only if $i \in a$, or equivalently if variable $\sigma_i$ participates in the interaction $\psi_a$. (b) The flow of messages contributing to the BP update of the message $M_{i \to a}^{t+1}(\sigma_i)$. (c) For the special case of binary Ising variables $\sigma_i$, the BP messages $M_{j \to b}^t(\sigma_j)$ and $M_{b \to i}^{t+1}(\sigma_i)$ can be parameterized in terms of the magnetizations $m_{j \to b}^t$ and $m_{b \to i}^{t+1}$. Furthermore, in the special case of pairwise interactions, $m_{b \to i}^{t+1}$ is solely a function of $m_{j \to b}^t$ (in addition to the coupling constant $J_{ij}$ and the temperature), so we can write the BP updates solely in terms of $m_{j \to b}^t$, which we rename to $m_{j \to i}^t$, which can be thought of as the magnetization of spin $\sigma_j$ in a cavity system with the coupling $J_{ij}$ removed. (d) The flow of messages contributing to the BP update of the cavity magnetization $m_{j \to i}^{t+1}$ in Eq. 3, again specialized to the case of Ising spins with pairwise interactions.

BP can then be viewed as an iterative dynamical algorithm for computing a marginal $P(\sigma_i)$ by passing messages along the factor graph. In the case of combinatorial optimization, we can focus on the zero temperature $\beta \to \infty$ limit. We will first describe the BP algorithm intuitively, and later give justification for it. BP employs two types of messages: one from variables to factors and another from factors to variables. Each message is a probability distribution over a single variable. We denote by $M_{j \to b}^t(\sigma_j)$ the message from variable $j$ to factor $b$ at iteration time $t$. It can be thought of intuitively as an approximation to the marginal distribution of $\sigma_j$ induced by all other interactions $a \neq b$. Thus $M_{j \to b}^t(\sigma_j)$ models the distribution over $\sigma_j$ in a cavity system in which interaction $b$ has been removed. Furthermore, we denote by $M_{b \to i}^t(\sigma_i)$ the message from factor $b$ to variable $i$ at iteration time $t$. Intuitively, we can think of $M_{b \to i}^t(\sigma_i)$ as the distribution on $\sigma_i$ induced by the direct influence of interaction $b$. These messages are called beliefs, as they indicate various probabilities that different interactions believe a single variable should assume. The BP equations amount to updating these beliefs to make them self-consistent with each other. For example, the (unnormalized) update equation for messages from factors to variables (see Fig. 7b) takes the form $M_{b \to i}^{t+1}(\sigma_i) = \sum_{\sigma_{b/j}} \psi_b(\sigma_b) \prod_{j \in b/i} M_{j \to b}^t(\sigma_j)$. Here, $b/i$ denotes the set of all variables in set $b$ with variable $i$ removed. Intuitively, the direct influence $M_{b \to i}^{t+1}(\sigma_i)$ of interaction $b$ on variable $i$ is computed by integrating out of the factor $\psi_b(\sigma_b)$ all variables $j$ participating in interaction $b$ other than $i$, supplemented by accounting for the effect of all other interactions besides $b$ by the product of beliefs $M_{j \to b}^t(\sigma_j)$ over all variables $j \in b/i$. This product structure essentially encodes an implicit assumption that the variables $j \in b$ would be independent of each other if interaction $b$ were removed. This would be exactly true if the factor graph were a tree with no loops. Similarly, the messages from variables to factors are updated via $M_{i \to a}^{t+1}(\sigma_i) = \prod_{b \in i/a} M_{b \to i}^{t+1}(\sigma_i)$ (see Fig. 7b). Intuitively, the belief $M_{i \to a}^{t+1}(\sigma_i)$ on variable $i$ induced by all other interactions besides $a$ is simply the product of the direct influences $M_{b \to i}^{t+1}(\sigma_i)$ of all interactions $b$ that involve variable $i$ besides interaction $a$ (this set of interactions is denoted by $i/a$). Belief propagation involves randomly initializing the messages and repeating these iterations until convergence. If the messages converge, the marginal distribution of any variable can be computed as $P(\sigma_i) = \prod_{a \in i} M_{a \to i}^t(\sigma_i)$. In essence the marginal is the product of all direct influences $M_{a \to i}^t(\sigma_i)$ over all interactions $a$ that contain variable $i$.

For a general factor graph, there is no guarantee that the BP update equations will converge in finite time, and even if they do, there is no guarantee the converged messages will yield accurate marginal distributions. However, if the factor graph is a tree, then it can be proven that the BP update equations do indeed converge, and moreover they converge to the correct marginals.[64] Moreover, even in graphs with loops, the fixed points of the BP update equations were shown to be in one to one correspondence with extrema of a certain Bethe free energy approximation to the true free energy associated with the factor graph distribution.[71] This observation yielded a seminal connection between BP in computer science, and the Bethe approximation in statistical physics. The exactness of BP on tree graphs, as well as the variational connection between BP and Bethe free energy on graphs with loops, motivated the further study of BP updates in sparsely connected random factor graphs in which loops are of size O(log N). In many such settings BP updates converge and yield good approximate marginals.[66] In particular, if correlations between variables $i \in a$ adjacent to a factor $a$ are weak upon removal of that factor, then BP is thought to work well.

Now specializing to the case of Ising spins in which each variable $\sigma = \pm 1$, every message $M(\sigma)$ is a distribution over a single spin, and can be uniquely characterized by a field $h$ via the relation $M(\sigma) \propto e^{\beta h}$ or equivalently through the magnetization $m = tanh(\beta h)$. Thus the BP update equations for the beliefs $M_{j \to b}^t(\sigma_j)$ and $M_{b \to i}^t(\sigma_i)$ can be viewed as a discrete time dynamical system on a collection of real valued fields $h_{j \to b}^t$ and $h_{b \to i}^t$ or equivalent magnetizations $m_{j \to b}^t$ and $m_{b \to i}^t$. Furthermore, in the case of pairwise interactions, where between any pair of spins there is at most one interaction, we can further simplify the BP equations. For instance, along any directed edge *from* spin $\sigma_j$ *to* spin $\sigma_i$ passing through an interaction $b$ with coupling constant $J_{ij}$, BP maintains two messages, parameterized by $m_{j \to b}^t$ and $m_{b \to i}^{t+1}$ where $m_{b \to i}^{t+1}$ is solely a function of $m_{j \to b}^t$, $J_{ij}$, and $\beta$ (Fig. 7(c)). Thus we can write the BP update equations for Ising systems solely in terms of one of the messages, which we rename to be $m_{j \to i}^t \equiv m_{j \to b}^t$. Thus for each connection in the Ising system, there are now two magnetizations: $m_{j \to i}^t$ and $m_{i \to j}^t$ corresponding to messages flowing along the two directions of the connection. Intuitively, $m_{j \to i}^t$ is the magnetization of spin $\sigma_j$ in a cavity system where the coupling $J_{ij}$ has been removed. Similarly, $m_{i \to j}^t$ is the magnetization of spin $\sigma_i$ in the same cavity system with coupling $J_{ij}$ removed. Some algebra reveals[66][68] that the BP equations in terms of the cavity magnetizations $m_{j \to i}^t$ are given by

$$m_{i \to j}^{t+1} = \tanh\left[\sum_{k \in i/j} \tanh^{-1}\{\tanh(\beta J_{ik}) m_{k \to i}^t\}\right]. \quad (3)$$

Here the sum over $k \in i/j$ denotes a sum over all neighbors of spin $i$ other than spin $j$. See Fig. 7d for a visualization of the flow of messages underlying Eq. 3. These BP equations maintain *two* magnetizations associated with each *connection* in the Ising system, corresponding to the two directions of flow across each edge. The messages are initialized randomly and updated according to equation (3), with the magnetizations thus flowing bi-directionally through the Ising network, hopefully converging to a set of fixed point magnetizations. The marginal of a spin $\sigma_i$ at iteration $t$ is then given by the magnetization $m_i^t = \tanh[\sum_{k \in i} \tanh^{-1}\{\tanh(\beta J_{ik}) m_{k \to i}^t\}]$. While this dynamics promotes the spin variables to analog variables, like the classical CIM dynamics in Eq. 1, it also bears three salient differences from the CIM dynamics: it operates in discrete time, maintains two analog variables per edge, rather than one analog variable per spin, and uses a different nonlinearity. Of course, the specialized BP dynamics are expected to work well specifically for classes of sparsely connected tree-like Ising systems in which removing an interaction $J_{ij}$ from the system makes the spins $\sigma_i$ and $\sigma_j$ approximately independent, but are not guaranteed to yield accurate marginals in more general settings.

The BP equations for Ising systems can also be used to derive the famous TAP equations[72] for the Sherrington Kirkpatrick (SK) model,[35] where each coupling constant $J_{ij}$ is chosen i.i.d from a zero mean Gaussian distribution with variance $1/N$. Because the couplings are now $O(1/\sqrt{N})$, we can Taylor expand Eq. 3 to obtain $m_{i \to j}^{t+1} = \tanh[\sum_{k \in i/j} \beta J_{ik} m_{k \to i}^t]$. Now this update equation is written in terms of cavity magnetizations $m_{i \to j}^{t+1}$ in a system in which a single coupling $J_{ij}$ is removed. Because the SK model connectivity is dense with each individual coupling $O(1/\sqrt{N})$, each cavity magnetization $m_{i \to j}^{t+1}$ with $J_{ij}$ removed, is close to the actual magnetization $m_i^{t+1}$ when $J_{ij}$ is present. By Taylor expanding in the small difference between $m_{i \to j}^{t+1}$ and $m_i^{t+1}$, one can write the BP update equations for the case of dense mean field connectivity solely in terms of the variables $m_i^{t+1}$ (see [68] for a derivation of the TAP equations from this BP perspective):

$$m_i^{t+1} = \tanh\left[\sum_j \beta J_{ij} m_j^t - \beta^2 m_i^{t-2} \sum_j J_{ij}^2 \{1 - (m_j^{t-1})^2\}\right] \quad (4)$$

This achieves a dramatic simplification in the dynamics of Eq. 3 from tracking $2N^2$ variables to only tracking N variables, and as such is more similar to the CIM dynamics in Eq. 1. Again there are still several differences: the dynamics in Eq. 4 is discrete time, uses a different nonlinearity, and has an interesting structured history dependence extending over two time steps. Remarkably, although BP was derived with the setting of sparse random graphs in mind, the particular form of the approximate BP equations for the dense mean field SK model can be proven to converge to the correct magnetizations as long as the SK model is outside of the spin glass phase.[73]

So far, we have seen a set of analog approaches to solving Ising systems in specialized cases

(sparse random and dense mean field connectivities). However, these local update rules do not work well when such connectivities exhibit spin glass behavior. It is thought that the key impediment to local algorithms working well in the spin glass regime is the existence of multiple minima in the free energy landscape over spin configurations.[66] This multiplicity yields a high reactivity of the spin system to the addition or flip of a single spin. For example, if a configuration is within a valley with low free energy, and one forces a single spin flip, this external force might slightly raise the energy of the current valley and lower the energy of another valley that is far away in spin configuration space but nearby in energy levels, thereby making these distant spin configurations preferable from an optimization perspective. In such a highly reactive situation, flipping one spin at a time will not enable one to jump from valleys that were optimal (lower energy) before the spin flip, to a far away valley that is now more optimal (even lower energy) after the spin flip. This physical picture of multiple valleys that are well separated in spin configuration space, but whose energies are near each other, and can therefore reshuffle their energy orders upon the flips of individual spins, motivated the invention of alternative algorithms that extend belief propagation to survey propagation. The key idea, in the context of an Ising system, is that the magnetizations $m_{i \to j}$ of BP now correspond to the magnetizations of spin configurations in a *single* free energy valley (still in a cavity system with the coupling $J_{ij}$ removed). SP goes beyond this to keep track of the *distribution* of BP messages *across* all the free energy valleys. We denote this distribution at iteration $t$ by $P^t(m_{i \to j})$. The distribution over BP beliefs is called a *survey*. SP propagates these surveys, or distributions over the BP messages across different valleys, taking into account changes in the free energy of the various valleys before and after the addition of a coupling $J_{ij}$. This more nonlocal SP algorithm can find solutions to hard constraint satisfaction problems in situations where the local BP algorithm fails.[65] Furthermore, recent work going beyond SP, but specialized to the SK model, yields message passing equations that can probably find near ground state spin configurations of the SK model (under certain widely believed assumptions about the geometry of the SK model's free energy landscape) but with a time that grows with the energy gap between the found solution and the ground state.[36]

Interestingly, the promotion of the analog magnetizations $m_{i \to j}^{t+1}$ of BP to distributions $P^t(m_{i \to j})$ over these magnetizations is qualitatively reminiscent of the promotion of the classical analog variables of the CIM to quantum wavefunctions over these variables. However this is merely an analogy to be used as a potential inspiration for both understanding and augmenting current quantum CIM dynamics. Moreover, the SP picture cannot account for quantum correlations. Overall, much further theoretical and empirical work needs to be done in obtaining a quantitative understanding the behavior of the CIM in the quantum regime, and the behavior of SP for diverse combinatorial Ising spin systems beyond the SK model, as well as potential relations between the two approaches. An intriguing possibility is that the quantum CIM dynamics enables a nonlocal parallel search over multiple free energy valleys in a manner that may be more powerful than the SP dynamics due to the quantum nature of the CIM.

Future Outlook

While current MFB-CIM hardware implementations would not seem capable of sustaining even limited transient entanglement because of their continual projection of each spin-amplitude on each round trip, it is possible that near-term prototypes could probe quantum-perturbed CIM dynamics at least in the small-$N$ regime. A recent analysis[74] of a modified MFB-CIM architecture utilizing entanglement swapping-type measurements shows that it should be possible to populate entangled states (of specific structure determined by the measurement configuration) of the spin-amplitudes, if the round-trip optical losses can be made sufficiently small. This type of setup could be used to enable certain entanglement structures to be created by transient non-local flow of quantum states through phase space, or to create specific entangled initial states for future CIM algorithms that exploit quantum interference in some more directed way. One may speculate that the impact of quantum phenomena could become more pronounced in CIMs with extremely low pump threshold, for which quantum uncertainties could potentially be larger relative to the scale of topological structures in the mean-field (in a quantum-optical sense) phase space in the critical near-threshold regime. Prospects for realizing such low-threshold CIM hardware have recently been boosted by progress towards the construction of optical parametric oscillators using dispersion-engineered nanophotonic lithium niobate waveguides and ultra-fast pump pulses.[75]

For methods that rely on the relaxation of a potential function, either a Lyapunov function for dynamical systems or free energy landscape for Monte Carlo simulations, it is generally believed that the exponential increase in the number of local minima is responsible for the difficulty in finding the ground-states. It has been suggested that the presence of an even greater number of critical points may prevent the dynamics from descending rapidly to lower energy states.[76] On the other hand, several recently proposed methods that rely on chaotic dynamics instead of a potential function have achieved good performance in solving hard combinatorial problems,[24][26][62][77][78][79] but the theoretical description of the number of non-trivial traps (limit-cycles or chaotic attractors) in their dynamics is lacking. It is of great interest to extend the study of complexity[76] (that is, the enumeration of local minima and critical points) to the case of chaotic dynamics for identifying the mechanisms that prevent these heuristics to find optimal solutions of combinatorial optimization problems and to derive convergence theorems and guarantees of returning solutions within a bounded ratio of the ground-state energy.

The closed-loop CIM has been proposed for improving the mapping of the Ising Hamiltonian when the time-evolution of the system is approximated to the first moment of the in-phase component distribution. Because the CIM has the potential of quantum parallel search[22] if dissipation can be reduced experimentally, it is important to extend the description of the closed-loop CIM to higher moments in order to identify possible computational benefits of squeezed or non-Gaussian states. In order to investigate this possibility but abstain from the difficulties of reaching a sufficiently low dissipation experimentally, the simulation of the CIM in digital hardware is necessary.

Another interesting prospect of the CIM is its extension to neuroscience research. One possibility is about merged quantum and neural computing concept. In the quantum theory of CIM, we start with a density operator master equation which takes into account a parametric gain, linear loss, gain saturation (or back conversion loss) and dissipative mutual coupling. By expanding the density operator with either a positive *P*-function (off-diagonal coherent state expansion), truncated Wigner-function or Husimi-function, we can obtain the quantum mechanical Fokker-Planck equations. Using the Ito rule in the Fokker-Planck equations, we finally derive the *c*-number stochastic differential equations (c-SDE).[27][28][49] We can use them for numerical simulation of the CIM on classical digital computers. This phase space method of quantum optics can be readily modified for numerical simulation of an open-dissipative classical neural network embedded in thermal reservoirs, where vacuum noise is replaced by thermal noise. We note that an ensemble average over many identical classical neural networks driven by independent thermal noise can reproduce the analog of quantum dynamics (entanglement and quantum discord) across bifurcation point. This scenario suggests a potential "quantum inspired computation" might be already implemented in the brain. Using the c-SDE of CIM as heuristic algorithm in classical neural network platform, we can perform a virtual quantum parallel search in cyber space. In order to compute the dynamic evolution of the density operator, we have to generate numerous trajectories by *c*-SDE. This can be done by ensemble averaging or time averaging.

However, what we need in the end is only the CIM final state, which is one of degenerate ground states, and in such a case, producing just one trajectory by c-SDE is enough. This is the unique advantage of the CIM approach and provided by the fact that this system starts computation as a quantum analog device and finishes it as a classical digital device. It is an interesting open question if the classical neural network in the brain implements such c-SDE dynamics driven by thermal reservoir noise. One of the important challenges in theoretical neuro-science is to answer how large number of neurons collectively interact to produce a macroscopic and emergent order such as decision making, cognition and consciousness via noise injected from thermal reservoirs and critical phenomena at phase transition point.[80][81][82][83] The quantum theory of the CIM may shed a light on this interesting frontier at physics and neuro-science interface.

Above we also reviewed a set of qualitative analogies connecting the CIM approach to combinatorial optimization with other approaches in computer science. In particular, we noted that just as the CIM dynamics involves a promotion of the original binary spin variables to classical analog variables and then quantum wave functions associated with these classical variables, computer science based approaches to combinatorial optimization also involve a promotion of the spin variables to analog variables (cavity magnetizations in BP for sparse random connectivities and magnetizations in TAP for dense mean field connectivities), and then distributions over magnetizations in SP. These analogies form a bridge between two previously separate strands of intellectual inquiry, and the cross-pollination of ideas between these strands could yield potential insights in both fields. In particular such cross-pollination may both advance the scientific understanding of and engineering improvements upon CIM dynamics.

For example, can convergence proofs of BP or TAP equations for special classes of Ising systems shed light on CIM dynamics in these same systems? Can differences between BP, TAP or SP dynamics and CIM dynamics suggest the rational design of better hardware modifications to the CIM, and would such modifications yield improved performance in Ising systems where BP, TAP and SP are known to converge? How would such modifications fare in more difficult settings where BP/TAP and even SP do not converge? Can the success of the CIM in problems other than random Ising systems for which BP, TAP and SP are specialized, suggest more general algorithms that work in other classes of Ising systems? Can the success of the CIM motivate other types of annealing schedules to the energy landscape that could serve as improvements on existing BP, TAP or SP algorithms? Overall the parallels and differences between BP, TAP, SP and CIM thus motivate many intriguing directions for future research that are as of yet completely unexplored.

Recently, a variety of different experimental platforms have been proposed and demonstrated for the implementation of Ising spin models.[84][85][86][87][88][89][90] It is expected that theoretical and experimental studies of coherent Ising machines mutually motivate and accelerate the advancement of the field synchronously.

More generally, we hope this article provides a sense of the rich possibilities for future interdisciplinary research focused around a multifaceted theoretical and experimental approach to combinatorial optimization uniting perspectives from statistics, computer science, statistical physics, and quantum optics, and making contact with diverse topics like dynamical systems theory, chaos, spin glasses, and belief and survey propagation.

ACKOWLEDGEMENTS

The authors wish to thank Z. Toroczkai, R. L. Byer, M. Fejer, B. Lev, A. Safavi-Naeini, A. Marandi, P. McMahon, D. Englund, W. Oliver, E. Rieffel, P. Ronagh, P. Drummond and M. Reid for their useful discussions.

DATA AVAILABILITY

The data that support the finding of this study are available from the corresponding author upon reasonable request.